\documentclass[runningheads]{svmult} 
\usepackage{subeqnar}  
\usepackage{multicol}  
\usepackage{cropmark} 
\usepackage{physprbb}  
\usepackage[dvips]{graphicx}
\usepackage{amsmath,amssymb}
\newcommand{\Mat}[1]{{{\boldsymbol{#1}}}}
\newcommand{\abs}[1]{\left\vert#1\right\vert}
\date{}
\begin{document}
\title*{\bf Gravitational Energy Loss and Binary Pulsars \\
in the Scalar Ether-Theory of Gravitation}
\titlerunning{Gravitational Energy Loss in a Scalar Theory of Gravity} 
\author{M. Arminjon\inst{}}
\institute{Laboratoire ``Sols, Solides, Structures'' [Unit\'e Mixte de Recherche of the CNRS], 
\\ BP 53, F-38041 Grenoble cedex 9, France.
\\ Email: arminjon@hmg.inpg.fr}
\maketitle
\begin{abstract} Motivation is given for trying a theory of gravity with a preferred reference frame (``ether'' for short). One such theory is summarized, that is a scalar bimetric theory. Dynamics is governed by an extension of Newton's second law. In the static case, geodesic motion is recovered together with Newton's attraction field. In the static spherical case, Schwarzschild's metric is got. An asymptotic scheme of post-Minkowskian (PM) approximation is built by associating a conceptual family of systems with the given weakly-gravitating system. It is more general than the post-Newtonian scheme in that the velocity may be comparable with $c$. This allows to justify why the 0PM approximation of the energy rate may be equated to the rate of the Newtonian energy, as is usually done. At the 0PM approximation of this theory, an isolated system loses energy by quadrupole radiation, without any monopole or dipole term. It seems plausible that the observations on binary pulsars (the pulse data) could be nicely fitted with a timing model based on this theory.
\end{abstract}

\section{Introduction}


The dominant opinion is that we already have both an excellent theory of gravitation -- namely, Einstein's general relativity (GR), and an excellent theoretical framework for particle physics -- namely, the standard model, which is based on quantum field theory (QFT). The important problem of making quantum theory and GR compatible together should then be solved essentially by going to a superstructure including the two former theories as particular and/or limiting cases, the candidates for that superstructure being searched in the wide range of string theory. This might prove to be the case in the future, but it must be admitted that string theories are very complex, being based on differential geometry on high-dimensional manifolds. Even GR and QFT separately are already very complex, to the point that the link between either theory and its experimentally confirmed predictions is not fully understood, being based on rather peculiar algorithms \cite{Feyerabend}. Therefore, apart from the main line of research, one can find some risky attempts, in which it is tried to take a quite different route. \\

One such route is suggested by the existence of an alternative version/ interpretation of special relativity, which version was initiated by Lorentz \cite{Lorentz} and Poincar\'e \cite{Poincare1905,Poincare1906}. According to this version, the ``relativistic'' effects, which essentially follow from the Lorentz transformation, are all due to the ``true'' Lorentz contraction of physical objects as they are moving through the ``ether'' or fundamental inertial frame \cite{Builder,Prokhovnik}. In fact it can be shown that both Lorentz's space-contraction and Larmor's time-dilation follow from the negative result of the Michelson-Morley experiment, under the assumption that light propagate in all directions with the same velocity $c$ with respect to a rigid reference frame, in which Euclidean geometry applies \cite{Prokhovnik,Arm93b}. In this ``Lorentz--Poincar\'e version of special relativity'', the time which is defined in an inertial frame by using the Poincar\'e-Einstein clock synchronization is generally not the ``true time'': only in the fundamental inertial frame does the Poincar\'e-Einstein-synchronized time coincide with the true time \cite{Poincare1900,Poincare1904,Prokhovnik}. The Lorentz--Poincar\'e version of special relativity has the same equations as the usual version initiated by Einstein in his well-known June 1905 paper, and is observationally equivalent to the usual version \cite{Prokhovnik}. However, just because of this, the ``ether'', which is still there in the former version, is indetectable. In particular, the true time of the Lorentz--Poincar\'e version is not experimentally accessible. \\

But, precisely, an essential assumption of the investigated alternative theory of gravitation \cite{Arm93a,Arm93b,Arm96a,Arm98b} is that the presence of a gravitational field breaks the Lorentz symmetry, thus making the ether detectable \cite{Arm96a}. Therefore, in this theory, the inertial time in the ether frame is a physically identifiable preferred time $T$ (called the ``absolute time''). It is accepted that the existence of a preferred space-time foliation could give a solution to extend quantum theory to the situation with gravitation \cite{Butterfield}. The very beginning of this program is the formulation of the free Klein-Gordon wave equation in a gravitational field, which is ambiguous in GR, but leads to a unique equation (which is a preferred-frame one) in the framework of the ``ether-theory'' \cite{Arm98a}. Another motivation for trying a very alternative theory of gravitation is to solve some problems which are common to GR and to most extensions of it -- namely, the existence of singularities (there is indeed no singularity in the investigated theory, neither during gravitational collapse ``in free fall'' with spherical symmetry \cite{Arm97} nor in homogeneous cosmological models \cite{Arm01a}), and the interpretation of the gauge condition (there is no gauge condition in this scalar theory).\\
	
This theory gives the same predictions as GR for light rays \cite{Arm98c} and it unambiguously predicts that the cosmic expansion must be accelerated \cite{Arm01a}. As to celestial mechanics: since this theory predicts Schwarzschild's motion for test particles in the static case with spherical symmetry \cite{Arm93b,Arm96a}, one a priori expects that it should improve celestial mechanics as compared with Newton's theory, in the same way as GR. However, due to the existence of preferred-frame effects, one has to check this carefully. To this end, an ``asymptotic'' scheme of post-Newtonian (PN) approximation has been built for a weakly self-gravitating system of extended bodies, that consists in associating a one-parameter family of systems with that given system: each system is defined by an initial-value problem \cite{Arm00a}. This is a correct way to obtain asymptotic expansions of the solution fields, with respect to the small weak-field parameter. For extended bodies in GR, this has been done only with a very particular initial condition as to the space metric \cite{FutaSchutz}, and the obtained local PN equations have not been integrated in the volume of the bodies to get equations of motion of the mass centers. The latter has been done for the investigated scalar theory \cite{Arm00b,Arm03b}. The obtained equations of motion of the mass centers depend on the spin rates of the bodies and involve structure-dependent parameters -- two natural features for a theory that integrates the mass-energy equivalence. In order to test the theory, these equations have been implemented \cite{Arm02d,Arm03c} in a code for numerical integration and parameter adjustment, that has been built by the author and tested \cite{Arm03a,Arm02a} with standard equations of motion. (For the moment, the spin rates of the main bodies of the solar system have been neglected.) It has thus been found that the difference with a standard ephemeris \cite{Standish95} does not exceed 3'' over the 20th century \cite{Arm03c}. One may expect to significantly improve the observational agreement by improving the numerical treatment and also by adjusting the equations, not on an ephemeris, but directly on observational data. Indeed some correction factors of observational data
are taken as free parameters in the adjustment of an ephemeris, hence the observations are not completely independent of the gravitational model. \\

Thus the current situation as to the experimental test of this theory is encouraging. However, it is usually considered that a scalar theory should have problems with gravitational radiation \cite{Carlip}. (Note that it is usually considered that a scalar theory should also have problems with gravitational effects on light rays  \cite{MTW}, and that it is not the case for the present preferred-frame theory.) Therefore, the aim of the present investigation was to examine gravitational radiation. Before discussing the main line of that investigation, we shall present the theory.

\section{Basic Principles of this Scalar Theory}\label{scalartheory}

\subsection{Space-time metric and field equation}\label{metric+field}

The theory considered is a {\it bimetric theory}: space-time is endowed with both a flat metric $\Mat{\gamma}^0$ and a curved metric $\Mat{\gamma}$. The relation between the two metrics is fixed by just a scalar field $f$ \cite{Arm93b}. This can be true only because this is a {\it preferred-frame theory} and the equations are written in the preferred reference frame $\mathrm{E}$ (``ether''). 
\footnote{\,The theory could be rewritten in a generally-covariant form by incorporating the velocity field of the ether (relative to the reference frame associated with the given coordinate system) into the fields of the theory. If one restricts the choice of the reference frame to the ``inertial'' class made of the frames that have a uniform rigid motion with respect to $\mathrm{E}$ (the uniformity and rigidity being defined in terms of the flat metric $\Mat{\gamma}^0$), then one just has to add a constant vector $\mathbf{V}$ to the unknowns.
} 
Thus, space-time is the product $\mathbf{R}\times \mathrm{M}$, where $\mathrm{M}$ is the ``space" manifold, i.e. the set of the positions $\mathbf{x}$ in the preferred reference frame $\mathrm{E}$, and most equations are merely space-covariant. The preferred time coordinate $T$ is the inertial time in the frame $\mathrm{E}$ (i.e. $T=x^0/c$ with $c$ the velocity of light, and where the coordinates $(x^\mu) (\mu =0,...,3)$ are Galilean for the metric $\Mat{\gamma}^0$, that is, $\gamma^0_{\mu \nu} = \eta_{\mu \nu}$ with $(\eta_{\mu \nu})=\mathrm{diag}(1,-1,-1,-1)$). In particular, the scalar field $f \equiv (\gamma_{00})_{\mathrm{E}}$ (thus in coordinates $(y^\mu)$ bound to the frame $\mathrm{E}$ and with $y^0=cT$) determines the slowing down of real clocks in a gravitational field, as compared with an ``absolute clock'' that would measure the absolute time $T$. This is the case for a clock that is fixed in the ether frame, and far enough from any massive body so that it is not affected by any gravitational field \cite{Arm93b}.\\

The equation for the field $f$ is \cite{Arm96a} a {\it nonlinear wave equation}: 
\begin{equation} \label{field}
\Delta f- \frac{1}{f} \left(\frac{f_{,0}}{f}\right)_{,0}= \frac{8\pi G}{c^2} \sigma \qquad (y^0 = cT),
\end{equation}
with $\Delta \equiv \mathrm{div}_{\Mat{g}^0} \mathrm{grad}_{\Mat{g}^0}$ the usual Laplace operator defined with the Euclidean metric $\Mat{g}^0$, which is the spatial part of the flat metric $\Mat{\gamma}^0$ in the frame $\mathrm{E}$; $G$ is Newton's gravitation constant and $\sigma \equiv (T^{00})_{\mathrm{E}}$ is the mass-energy density in the ether frame, $\mathbf{T}$ being the ``mass tensor'', i.e. the energy-momentum tensor in mass units. In the most general case, applicable to a heterogeneous universe on a cosmological time-scale, the gravitational field is not directly the field $f$ but instead the field of the ``ether pressure'' $p_e$, and a different field equation applies \cite{Arm93b}. But this reduces to (\ref{field}) if the time variation of the ``reference pressure'' $p_e^\infty (T)\equiv \mathrm{Sup}_{\mathbf{x} \in\mathrm{M}} p_e(\mathbf{x},T)$ is neglected \cite{Arm96a}, which should be the case except on a cosmological time-scale. The analysis of homogeneous cosmological models \cite {Arm01a} confirms that the variation of $p_e^\infty (T)$ takes place over long time scales, of the order of $10^8$ years.

\subsection{Dynamical equations}\label{dynamics}

Motion of a {\it free test particle} is defined by an extension of the special-relativistic form of Newton's second law:\\
\begin{equation} \label{Newtonlaw}
\frac{E}{c^2}\,\mathbf{g} = \frac{D\mathbf{P}}{Dt_\mathbf{x}},
\end{equation}
where $E$ is the energy of the test particle: for a mass particle, $E=m(v)c^2$ with $m(v)=\gamma_v m(0)$ the velocity-dependent inertial mass ($\gamma_v$ is the Lorentz factor); for a photon, $E = h\nu$ with $\nu$ the frequency measured with real clocks affected by the gravitational slowing down and thus measuring the ``local time'' $t_{\mathbf{x}}$ (see Eq. (\ref{velocity}) below). Moreover, 
\begin{equation}\label{gravity_accel}
\mathbf{g}=-\frac{c^2}{2}\nabla f 
\end{equation}
is the gravity acceleration,
\begin{equation}\label{velocity}
\mathbf{v}\equiv \frac{\D \mathbf{x}}{\D t_{\mathbf{x}}} \equiv \frac{1}{\sqrt{f}} \frac{\D \mathbf{x}}{\D T}
\end{equation} 
is the velocity, and $D/Dt_\mathbf{x}$ is the correct time-derivative of a vector in the space manifold $\mathrm{M}$ endowed with the time-dependent Riemannian metric $\Mat{g}$ ($\Mat{g}$ is the spatial part of $\Mat{\gamma}$ in the frame $\mathrm{E}$) \cite{Arm96a}. Finally, 
\begin{equation}\label{momentum}
\mathbf{P} \equiv \frac{E}{c^2}\,\mathbf{v} 
\end{equation} 
is the momentum of the test particle.\\

In the static case ($f_{,0}=0$), the extension~(\ref{Newtonlaw}) of Newton's second law (thus three scalar equations) implies that free test particles follow the geodesic lines of metric $\Mat{\gamma}$ (thus four scalar equations: the time component of the geodesic equation is equivalent to the energy equation valid for a test particle in a static field) \cite{Arm93b}. In the static case with spherical symmetry, Schwarzschild's exterior metric is obtained \cite{Arm93b}.\\

The dynamical equation for a {\it continuous medium} (fluid, electromagnetic field, ...) is deduced from Newton's second law as defined above for a test particle. Indeed, for a {\it dust}, Eq.~(\ref{Newtonlaw}) may be applied pointwise and implies \cite{Arm98b}:
\begin{equation} \label{continuum}
T_{\mu;\nu}^{\nu} = b_{\mu},				          
\end{equation}
\begin{equation} \label{definition_b}
b_0(\mathbf{T}) \equiv \frac{1}{2}\,g_{jk,0}\,T^{jk}, \quad b_i(\mathbf{T}) \equiv -\frac{1}{2}\,g_{ik,0}\,T^{0k} 
\end{equation}
(recall that $\Mat{g} = (g_{ij})$ is the (curved) space metric in the frame $\mathrm{E}$. Semicolon means, as usual, covariant differentiation defined with the connection associated to the (curved) space-time metric $\Mat{\gamma}$. Indices are raised and lowered with $\Mat{\gamma}$). The universality of gravity means that {\it the same equation must hold true for any continuous medium}.

\subsection{General energy conservation in this theory} 

Equation~(\ref{continuum}) may be rewritten in terms of the {\it flat} metric $\Mat{\gamma}^0$, giving as the time component:
\begin{equation} \label{flat_0}
T_{0,\nu}^\nu =  \frac{1}{2}f_{,0}\,T^{00}
\end{equation}
(valid in Galilean coordinates for the flat metric) \cite{Arm98b}. Using the equation (\ref{field}) for the scalar field $f$, the r.h.s. of Eq.~(\ref{flat_0}) may be transformed to a 4-divergence (with respect to the {\it flat} metric), thus giving a true conservation equation for the energy density
\begin{equation} \label{energydensity}
\varepsilon \equiv c^2 T_0^0 + \frac{1}{8\pi G} \left[\mathbf{g}^2 + \frac{c^4}{4}\left(\frac{f_{,0}}{f}\right)^2\right] \equiv \varepsilon_\mathrm{m} + \varepsilon_\mathrm{g}.
\end{equation}
\cite{Arm96a}. The energy rate in a fixed domain $\Omega$, whose boundary $\partial \Omega$ is not crossed by any matter, is got by integrating the conservation equation, it is
\begin{equation} \label{energyrate}
\dot{E} = -\Phi , \quad \Phi \equiv \frac{c^2}{8\pi G} \int_{\partial \Omega}\, \partial_T f \,\,\mathbf{g.n}\,\D S, \quad E \equiv \int_{\Omega}\, \varepsilon\,\D V.
\end{equation}
\\

\section{Gravitational Radiation}

In GR and other relativistic theories of gravitation, gravitational radiation is mostly studied with the ``weak-field approximation'', which is essentially a standard linearization of the field equations \cite{Fock64,L&L,Weinberg,MTW}. However, we are here in a domain where the accuracy of the observations (of the timing of the pulses emitted by binary pulsars) is very high, and so also is the precision of the comparison theory/ observation \cite{TaylorWeisberg}. Therefore, it is desirable that the link between the equations of the theory and its numerical predictions should be obtained with the help of a well-defined approximation scheme. Thus one would need to introduce an asymptotic framework, associating with the physical system of interest $\mathrm{S}$ a family ($\mathrm{S}_\lambda$) of gravitating systems ($\lambda$ is the small field-strength parameter). A such ``asymptotic scheme'' was previously developed for celestial mechanics (see the fourth paragraph of the introduction here). As $\lambda \rightarrow 0$, the behaviour becomes Newtonian. Thus it is a {\it post-Newtonian} (PN) scheme. Like the standard PN scheme \cite{Fock64,Chandra65,Weinberg,Will}, that ``asymptotic'' PN scheme leads to Poisson equations with instantaneous propagation, therefore it does not fit in the present context, since the aim is to describe how the energy balance is affected by the propagation of gravitational waves at the finite velocity $c$. \\

What one would need is an asymptotic {\it post-Minkowskian} (PM) scheme, i.e., one leading to a {\it wave equation} for the gravitational field (here the scalar $f$). In GR, it has been proved that one-parameter families of solutions to the vacuum Einstein equations generically exist, and that the successive coefficients of their Taylor expansions with respect to the parameter do satisfy the successive equations of the PM approximation \cite{DamourSchmidt}. However, in order to justify standard calculations based on equating the Newtonian energy rate to the energy rate as given by the ``quadrupole formula of GR'' (see e.g. Refs. \cite{L&L,Weinberg,Will}), it is necessary that there is indeed matter which has the relevant quadrupole tensor. One possible approach that would provide a reasonable justification to such calculations would be to show that the same given self-gravitating system may be consistently envisaged both in well-defined PN and PM approximation schemes; that the PM scheme (as applied to the given system and thus also inside the bodies) gives some calculable non-zero gravitational energy rate $\dot{E}_{\mathrm{PM}}$; and that the PM scheme is more general than the PN scheme, so that it does make sense to state the equation 
\begin{equation}\label{energy_rate_gross}
\dot{E} \simeq \dot{E}_{\mathrm{PM}} \simeq \dot{E}_{\mathrm{PN}}. 
\end{equation}
This turns out to be feasible. We shall insist on the essential points of that approach (the details of which shall be published elsewhere) and shall skip the calculations.

\subsection{Principle of the asymptotic PM approximation} 

As for the ``asymptotic'' PN scheme, which has been presented in detail in Ref. \cite{Arm00a}, a family ($\mathrm{S}_\lambda$) of perfect-fluid systems is defined by a family of {\it initial conditions}\,: at the initial time, the fields of pressure and proper rest-mass density in system $\mathrm{S}_\lambda$ have respectively the form
\begin{equation}\label{PM_IC1}
p^{(\lambda)}(\mathbf{x})=\lambda^2 p^{(1)}(\mathbf{x}), \quad
\rho^{(\lambda)}(\mathbf{x})=\lambda\rho^{(1)}(\mathbf{x});
\end{equation}
the initial condition for the gravitational field $f$, or rather for $V\equiv (c^2/2)(1-f)$ (which plays the role of the Newtonian potential), is given by
\begin{equation}\label{PM_IC2}
V^{(\lambda)}(\mathbf{x})=\lambda V^{(1)}(\mathbf{x}), \quad\partial_TV^{(\lambda)}(\mathbf{x})=\lambda \partial _TV^{(1)}(\mathbf{x});
\end {equation}
and the initial condition for the velocity is
\begin{equation}\label{PM_IC3}
\underline{\mathbf{u}^{(\lambda)}(\mathbf{x})= \mathbf{u}^{(1)}(\mathbf{x})}.
\end {equation}\\

The system of interest, $\mathrm{S}$ (e.g. a binary pulsar), is assumed to correspond to a small value $\lambda_0 \ll 1$ of the field-strength parameter 
\begin{equation}\label{deflambda}
  \lambda= \mathrm{Sup}_{\mathbf{x}\in \mathrm{M}}[1-f^{(\lambda)}(\mathbf{x},T=0)]/2 = \mathrm{Sup}_{\mathbf{x}\in \mathrm{M}}[V^{(\lambda)}(\mathbf{x},T=0)]/c^2.
\end{equation}
Hence, equations (\ref{PM_IC1})$_{1-2}$ and (\ref{PM_IC2})$_1$ mean physically that, in the system $\mathrm{S}$  (and more precisely near the center of the most massive body of $\mathrm{S}$), the scalar $1-f=1-f^{(\lambda_0)}$, that determines the magnitude of the difference between the flat and curved metrics, is of the order of magnitude $2\lambda_0$, and that the fields $p$ and $\rho$ have the order $\lambda_0^2$ and $\lambda_0$, respectively (in units such that the fields $p^{(1)}$ and $\rho^{(1)}$ are of the order of unity near the center of the most massive body). These equations (\ref{PM_IC1})$_{1-2}$ and (\ref{PM_IC2})$_1$ are exactly the same as in the PN scheme \cite{Arm00a}.
\footnote
{~The small parameter considered in the present paper is $\lambda=\varepsilon^2$, where $\varepsilon$  is
that used in Ref. \cite{Arm00a}.
}   
The essential difference with the (asymptotic) PN scheme of the scalar theory \cite{Arm00a} is that, in the PM scheme, it is not {\it necessary} that $\abs{\mathbf{u}} \ll c$ (in the physical sense) for the given system $\mathrm{S}$, because the velocity in the generic system $\mathrm{S}_\lambda$ of the family is order zero with respect to the parameter $\lambda$ (Eq.~(\ref{PM_IC3})) -- whereas the velocity is order $\sqrt{\lambda}$ in the PN scheme \cite{Arm00a}. Correspondingly, the time derivative of the gravitational potential $V$ is order 1 in $\lambda$ in the PM scheme (Eq.~(\ref{PM_IC2})$_2$), whereas it is order 3/2 in the PN scheme \cite{Arm00a}. Hence, the (asymptotic) PM scheme is just {\it more general} than the corresponding PN scheme, in the sense that the given system $\mathrm{S}$ will automatically be suitable for the PM scheme if it is for the PN one, and the converse is not true. (But the PM {\it family} ($\mathrm{S}_\lambda$) is not ``more general'' than the corresponding PN family: it is just a different family of solutions.) Thus, $\dot{E}_{\mathrm{PM}}$ can be used in the PN scheme.

\subsection {PM approximation of the energy rate}

We assume that the initial-value problem $\mathrm{P}_\lambda$, defined by Eqs.~(\ref{PM_IC1})--(\ref{PM_IC3}), has generically a unique solution in some large-enough domain, and that the solution fields admit asymptotic expansions in powers of $\lambda$, which have necessarily the same dominant order in $\lambda$ as the initial values. Precisely, we need just the principal part in these expansions. After changing the mass and the time units, for system $\mathrm{S}_\lambda$, to the new units $[\mathrm{M}]_\lambda = \lambda[\mathrm{M}]$ and $[\mathrm{T}]_\lambda = [\mathrm{T}]/\sqrt{\lambda}$ (where $[\mathrm{M}]$ and $[\mathrm{T}]$ are the starting units), the small parameter $\lambda$ becomes proportional to $1/c^2$, and all fields: $p$, $\rho$, $V$, $\partial V/\partial x^0$, {\it but the velocity}, are order zero, whereas the velocity field $\mathbf{u}$ is order $c$. We may hence write:
\begin{equation}\label{expans_fields}
V = U + O(1/c^2),\qquad \sigma = \sigma_0 + O(1/c^2),\qquad \mathbf{u} = \mathbf{u}_0c + O(1/c)
\end{equation}
(recall that $\sigma \equiv (T^{00})_{\mathrm{E}}$ is the source of the gravitational field, Eq.~(\ref{field})). These expansions have primarily to be valid at fixed values of the relevant time and space variables. Due to the fact that the velocity is order zero in the small parameter (in fixed units) for the PM family, the relevant time variable (the ``dynamical time'' in the terminology of Ref. \cite{FutaSchutz}) is the true time $T$ (as expressed in fixed units) -- whereas the relevant time variable is $T\sqrt{\lambda}$ for the PN family of systems \cite{Arm00a}. Therefore, in the new (varying) units, in which the expansions have the convenient form (\ref{expans_fields}), and in which $\lambda\propto 1/c^2$, the relevant time variable is $x^0 = cT$. As a consequence, the expansion of the field equation (\ref{field}) leads, as we wished, to the wave equation for the 0PM potential $U$:
\begin{equation}\label{expans_field_PM0}
\square U \equiv U_{,0,0}-\Delta U = 4\pi G \sigma_0 \qquad (x^0 = cT)			  
\end{equation}
 -- instead of the Poisson equation which applies to the 0PN (Newtonian) potential. As $\lambda \rightarrow 0$, the gravitational energy flux (\ref{energyrate}) has a principal part $\Phi_{0\mathrm{PM}}$:
\begin{equation}\label{PM0_flux}
\Phi = \Phi_{0\mathrm{PM}}\left(1+O( \lambda )\right), \quad \Phi_{0\mathrm{PM}} \equiv \frac{-1}{4\pi G} \int_{\partial \Omega} \partial_T U\,U_{,k}n^k\,\D S, 
\end{equation}
where $U$ is the relevant (retarded) solution of Eq.~(\ref{expans_field_PM0}).

\subsection {There are only quadrupole terms in the limit radiative flux}

 The next task is to calculate the limit, as $R \rightarrow \infty$, of the 0PM flux $\Phi_{0\mathrm{PM}}$ on the sphere $\partial \Omega : \abs{\mathbf{X}} = R$. To this end, one first expands the retarded potential $U$ as $R \rightarrow \infty$, then one inserts this in (\ref{PM0_flux}). The multipole expansion of the retarded potential $U$ contains monopole, dipole, and quadrupole terms, i.e. it contains
\begin{eqnarray}\label{def_M_d_J}
M\equiv \int \sigma_{0}  \D V, & \quad d_i(\mathbf{X},T) \equiv \int x^i\sigma_{0} (\mathbf{x},cT-\abs{\mathbf{X}}) \D V(\mathbf{x}),\\ & J_{ij}(\mathbf{X},T) \equiv \int x^ix^j\sigma_{0} (\mathbf{x},cT-\abs{\mathbf{X}}) \D V(\mathbf{x}). 
\end{eqnarray}
Indeed, we find:
\begin{equation}\label{expans_U_ret}
\frac{U(\mathbf{X},T)}{G}=\frac{M}{R} + \frac{X^i\dot{d}_i}{cR^2} + \frac{X^iX^j}{2c^2R^3}\ddot{J}_{ij} +...+ O\left(\frac{1}{R^2}\right),
\end{equation}
where overdot means partial derivative with respect to $T$ and the points of suspension indicate omitted terms of order $1/R$ coming from time derivatives of order $\geq 3$ in the Taylor series expansion of the retarded density. However, $M=\mathrm{Const.}$ disappears from the flux (\ref{PM0_flux}), due to the presence of the partial derivative $\partial_T U$. As to $d_i$, it is in the {\it limit} flux just by $\ddot{d}_i$. But $\ddot{d}_i=0$ from the 0PM conservation of the total momentum (which is obtained when inserting the expansions (\ref{expans_fields}) into the dynamical equation (\ref{continuum})). It follows that the limit as $R \rightarrow \infty $ of the PM energy flux contains just quadrupole terms:
\begin{equation}\label{limit_flux2}
(\Phi_{0\mathrm{PM}})_{\mathrm{lim}} = \frac{G}{60c^5} \left(2\,\mathrm{tr}\,\dot{\ddot{\mathbf{J}}}^2 + \mathrm{tr}^2 \dot{\ddot{\mathbf{J}}}\right).
\end{equation}

\subsection {Use of the PM approximation of the energy flux}

As already mentioned, the fact that the PM scheme is more general than the PN scheme implies that it makes sense to use the energy rate $\dot{E}_{\mathrm{PM}}$ in the PN scheme. Let us try to expand on this point. The limit, as $R \rightarrow \infty $, of the principal part of the PM energy flux, is given by Eq.(\ref{limit_flux2}), whereas, {\it at the zero-order PN approximation}, the corresponding limit is {\it zero}. Indeed, the principal part of the PN energy flux, $\Phi_{0\mathrm{PN}}$, is just the Newtonian energy flux. But the Newtonian energy of an isolated system is conserved, hence we have
\footnote{
This is easy to check by integrating the local balance equations of Newtonian gravity \cite{Arm96a}; the point is that, in Newtonian gravity, the gravitational energy flux has just the same expression (\ref{PM0_flux})$_2$
as in the 0PM approximation, but with $U$ being now the Newtonian potential, which is such that $U_{,k} = O(1/R^2)$ and $\partial_T U \rightarrow 0$ at large $R$.
}
\begin{equation}\label{Newt_limit_flux}
 \lim_{R \rightarrow \infty} \Phi_{0\mathrm{PN}} = 0.
\end{equation}
However, although the zero-order PN (i.e., Newtonian) energy $E_{0\mathrm{PN}}$ is thus constant at this same zero-order PN approximation, it has no reason to remain so at the following, truly post-Newtonian, approximations. Indeed, as the order $n$ of the PN approximation increases, the energy rate $\dot{E}^{(n)}_{\mathrm{PN}}$ calculated at the $n^{\mathrm{th}}$ PN approximation should become a better and better approximation of the exact energy rate $\dot{E}$, more precisely one should have
\begin{equation}\label{energy_rate_nPN}
\dot{E}= \dot{E}^{(n)}_{\mathrm{PN}} + O(\lambda^{n+1}). 
\end{equation}
Moreover, the dominant part of $\dot{E}^{(n)}_{\mathrm{PN}}$ should precisely be the rate of $E_{0\mathrm{PN}}$: if the PN expansions are uniform with respect to time, then, by summing the rates of the successive terms in the asymptotic expansion of the energy: 
\begin{equation}\label{energy_nPN_sums}
E = E_{\mathrm{0PN}} + \lambda (\delta E_{\mathrm{1PN}})+...+ O(\lambda^{n+1}), 
\end{equation}
one should obtain the asymptotic expansion of the energy rate:
\begin{equation}\label{energy_rate_nPN_sums}
\dot{E}= \frac{\D }{\D T}(E_{\mathrm{0PN}}) + \lambda \frac{\D }{\D T}(\delta E_{\mathrm{1PN}})+...+ O(\lambda^{n+1}). 
\end{equation}
More exactly, when we write Eq. (\ref{energy_rate_nPN_sums}), we are using the $n^{\mathrm{th}}$ PN approximation and we must hence rewrite this equation as
\begin{equation}\label{energy_rate_nPN_sums2}
\dot{E} = \dot{E}^{(n)}_{\mathrm{PN}} + O(\lambda^{n+1}) = \left(\frac{\D }{\D T}\right)_{n\mathrm{PN}} (E_{\mathrm{0PN}}) + \lambda \left(\frac{\D }{\D T}\right)_{n\mathrm{PN}} (\delta E_{\mathrm{1PN}})+...+ O(\lambda^{n+1}). 
\end{equation} 
Of course, just the same may be done with the PM scheme. What happens is that the first iterations of the {\it PN} scheme, in particular the very first one (the Newtonian approximation), give the simple prediction that the energy of the gravitating system is conserved, i.e. 
\begin{equation}\label{energy_rate=0}
\dot{E}^{(n)}_{\mathrm{PN}}  = 0 \quad \mathrm{for}\quad 0 \leq n < p.
\end{equation} 
Therefore, the exact energy rate is certainly very small, and one would like to calculate it by using the $p^{\,\,\mathrm{th}}$ PN approximation, that gives the first non-zero result (in GR, the standard PN scheme gives $p=5/2$ \cite{Chandra70}, the fractional number meaning merely that a Taylor expansion in $\lambda$ does not apply any more). But this is very complicated, hence one is led to admit that the first non-zero result given by the {\it PM} approximation, and which turns out to correspond to the zero-order PM approximation (Eq.~(\ref{limit_flux2})), is a good approximation to the exact result. The reason to believe so is that the PM approximation, in contrast to the PN one, does take into account naturally the phenomenon of gravitational radiation. We may summarize by admitting that, for some number $p$ (may be $p=5/2$):
\begin{equation} \label{energy_rate_ccl}
\dot{E} \simeq \dot{E}^{(p)}_{\mathrm{PN}} \simeq  \left(\frac{\D }{\D T}\right)_{p\mathrm{PN}} (E_{\mathrm{0PN}}) \simeq \left(\frac{\D }{\D T}\right)_{0\mathrm{PM}} (E_{\mathrm{0PM}}). 
\end{equation}

\subsection{Peters-Mathews coefficients and decrease in the orbital period}

Now we consider a binary system and we assume i) that the two bodies are far enough, so that they do not tidally interact; and ii) that the radiative energy loss does not affect the spins. Under these assumptions, the rates of the quadrupole tensor $\mathbf{J}$, which allow to approximate the gravitational radiative energy loss (Eq.~(\ref{limit_flux2})), may be calculated with the system being simply modelled as two Newtonian point masses. Then we calculate that 
\begin{equation}\label{binaryEdotETG}
\dot{E}\simeq - (\Phi_{0\mathrm{PM}})_{\mathrm{lim}}= -\frac{8}{15}\,\frac{G}{c^5} \,\frac{K^2}{r^4}\left[2\,\mathbf{u}^2-\frac{3}{4}\left(\mathbf{u.n}\right)^2\right], 
\end{equation}
where
\begin{equation}\label{def_x_u}
\mathbf{x}\equiv \mathbf{x}_1-\mathbf{x}_2, \, \mathbf{u}\equiv \dot{\mathbf{x}},\, K \equiv Gm_1m_2, \, r\equiv \abs{\mathbf{x}},\,\mathbf{n}\equiv \mathbf{x}/r
\end{equation}
($m_1$ and $m_2$ are the zero-order masses of the pulsar and its companion; $\mathbf{x}_1$ and $\mathbf{x}_2$ are their positions). Therefore, the Peters-Mathews coefficients are:
\begin{equation}\label{PetersMathewsETG}
k_1 = 2,\quad k_2=\frac{3}{4}, \quad k_\mathrm{dipole}=0 \qquad \mathrm{(scalar\,\,theory)}. 
\end{equation}
In GR in harmonic coordinates, the expression of the energy rate is \cite{Will}:
\begin{equation}\label{binaryEdotGR}
\dot{E}\simeq - (\Phi_{0\mathrm{PM}})_{\mathrm{lim}}= -\frac{8}{15}\,\frac{G}{c^5} \,\frac{K^2}{r^4}\left[12\,\mathbf{u}^2- 11\left(\mathbf{u.n}\right)^2\right]. 
\end{equation}
(I.e., the Peters-Mathews coefficients are in GR:
\begin{equation}\label{PetersMathewsGR}
k_1 = 12,\quad k_2=11, \quad k_\mathrm{dipole}=0 \qquad \mathrm{(GR\,\,in\,\,harmonic\,\,coordinates).)} 
\end{equation}
\\
In order to relate the energy rate to the rate of the orbital period $P$ of the binary, one uses Kepler's third law,
\begin{equation}\label{Kepler3}
(2\pi/P)^2 a^3 = G (m_1 + m_2), 
\end{equation}
and the expression of the Newtonian energy of the binary, 
\begin{equation}\label{Newtonianenergy}
E = - \frac{G m_1 m_2}{2a}
\end{equation}
(where $a$ is the semi-major axis of the relative orbit), and gets
\begin{equation}\label{period rate}
\frac{\dot{P}}{P} = -\frac{3}{2}\frac{\dot{E}}{E}.
\end{equation}
For binary pulsars, usually the ``companion" star is not seen and the observational input is the list of the arrival times of the successive pulses emitted by the pulsar, in short ``the pulse data'' \cite{TaylorWeisberg}. Essentially, these data show a periodic variation due to the motion of the pulsar around the mass center of the binary system, but there is a slow drift in this variation, due to the decrease in the period $P$. If one wants to test a given theory of gravitation, then the ``timing model'' used to fit the pulse data should be consistently and entirely based on that theory. If we apply Eqs.~(\ref{binaryEdotETG}) and~(\ref{period rate}) to a given binary pulsar, in taking all orbital parameters as they are obtained from a timing model based on GR in harmonic coordinates, then we will obtain a period change of the same order of magnitude as that found with the timing model based on GR, but significantly different from it -- because the energy rate (\ref{binaryEdotETG}) differs from that in GR (Eq.~(\ref{binaryEdotGR})), while keeping the same order of magnitude for a given orbit and a given $K$. In view of the foregoing sentence about the consistency of the timing model and the theory, this does not prove that the scalar theory is not able to accurately fit the pulse data (which is indeed the true test). It is worth to recall that, in contrast with this, most other alternative theories give a very different energy rate from that predicted by GR, usually they even predict the wrong sign \cite{TaylorWeisberg,Will}.

\section {Conclusion}

A scalar bimetric theory with a preferred reference frame has been motivated and summarized. An asymptotic framework has been introduced for the ``post-Minkowskian'' (PM) approximation outside {\it and inside} the bodies as well, by considering a family of initial-value problems for a perfect-fluid system; this seems to be new and could be useful in other theories. In particular, it has been shown that this PM scheme is more general than the corresponding post-Newtonian (PN) scheme and that, therefore, a simple and reasonable justification can be given for the procedure usually adopted in studies on gravitational radiation. That is, it is allowable to take the energy rate as given by the PM approximation and to say that it represents (approximately) the rate of the Newtonian energy. Using that PM scheme, the 0PM approximation of the energy rate for an isolated system has been calculated for the investigated scalar theory. This is indeed an energy loss, and it contains no dipole term (and no monopole term either). Thus, the radiative energy loss has the same structure in this theory as in GR, it has also the same order of magnitude. This is usually not the case for alternative theories, even for those that look much closer to GR. The author feels it plausible that this simple scalar theory might be able to fit accurately the pulse data of binary pulsars.

\subsection*{Note} 
The founding papers of special relativity, by Lorentz, Poincar\'e, Einstein, and Minkowski, can be found online on the web. Links may be found at\\
 \verb+http://geo.hmg.inpg.fr/arminjon/INTRO.html+.

\bibliographystyle{amsplain}

\end{document}